# Localized State-Induced Enhanced Intrinsic Phonon-Free Optical Transition in Silicon Nanocrystals

Feilong Wang (王飞龙), Qiongrong Ou (区琼荣) and Shuyu Zhang (张树宇) [*]

*Institute for Electric Light Sources, School of Information Science and Technology, Fudan University, Shanghai 200433, China PR*

[*] zhangshuyu@fudan.edu.cn

**ABSTRACT**

Silicon photoluminescence and lasing have been critical issues to breakthrough bottlenecks in the understanding of luminescence mechanisms. Unfortunately, long-standing disputes about the exciton recombination mechanism and fluorescence lifetime remain unresolved, especially about whether silicon nanocrystals (Si NCs) can realize fast direct-bandgap-like optical transitions. Here, using ground-state and excited-state density functional theory (DFT), we obtained intrinsic phonon-free optical transitions at sizes from $Si_{22}$ to $Si_{705}$, showing that very small Si NCs can realize a strong direct optical transition. Orbital labeling results show that this rapid transition does not come from the $\mathbf{\Gamma}$-$\mathbf{\Gamma}$-like transition, contrary to the conclusions from the effective mass approximation (EMA) and that $\mathbf{\Gamma}$-$\mathbf{X}$ mixing leads to a quasi-direct bandgap. This anomalous transition is particularly intense with decreasing size (or enhancement of quantum confinement). By investigating electron and hole distributions generated in the optical transition, localized state-induced enhanced emission (LIEE) in Si NCs was proposed. Quantum confinement (QC) distorts the excited-state electron spatial distribution by localizing Bloch waves into the NC core, resulting in increased hole and electron overlap, thus inducing a fast optical process. This work resolves important debates and proposes LIEE to explain the anomalous luminescence—a phase transition from weak (or none) luminescent state to strong optical transition, which will aid attempts at realizing high-radiative-rate NCs materials and application-level Si lasers.

## I.  INTRODUCTION

Silicon has dominated the microelectronics industry and shows potential larger-scale applications in waveguides [1,2], modulators [3,4], photodetectors [5], wavelength converters [6,7], etc. However, bulk silicon displays poor photoluminescent properties due to an indirect bandgap-inducing multiphonon-assisted radiative recombination [8], nonradiative recombination by defects [9] and free-carrier absorption [10], thus limiting its optical applications. No high-efficiency silicon light source has been manufactured thus far.

Si NCs provide prospects for silicon-based light sources [11–13], especially on-chip lasers earmarked for silicon-based optical interconnections. [14,15] Recently, a femtosecond-laser pumped Si NCs laser with an extremely low exciton average occupancy threshold by improved optical gain with high-pressure hydrogen passivation and suppressed Auger recombination using interface-state charge storage was manufactured [15–17], which indicates the feasibility of making a silicon laser at room temperature. However, the lasing spot intensity was limited, which can be attributed to long-life (and thus slow-rate) radiative recombination in Si NCs, limiting its important role in optical interconnection. Even if the fluorescence quantum efficiency of Si NCs can reach approximately 80% [18,19], the number of photons radiated per second is still 0.1~0.01% that of typical laser materials. [20,21]

Conversion of the indirect bandgap to the direct bandgap can meet this challenge. [22,23] In recent years, despite some limitations, some main theories have been proposed to rethink key problems in Si luminescence from the perspective of the energy band and greatly promote the development of Si NCs:

1. As the earliest EMA model was employed, A. A. Prokofiev *et al.* inferred that Si NCs may realize a direct transition since two points in the $\Gamma$-valence band with opposite effective mass would separate and produce a redshift with a decrease in size, and this unsteady, fast direct transition (F-band) was observed experimentally. [24] Correspondingly, de Boer *et al.* perfected the F-band experiment under different particle sizes and then determined that the fast transition would redshift with NCs size reduction. This means that QC and F-band may lead to a phase transition that makes Si NCs realize a direct bandgap when the size decreases to approximately 2 nm. [25] Nevertheless, the accurate atomic pseudopotential method overturned the negative effective mass, which was the cornerstone of the EMA [26] (**Fig. S1**). Nevertheless, no research has reported direct-transition Si NCs with the "redshift" after years of development. In fact, the redshift or blueshift of the $\Gamma$-$\Gamma$ transition due to size change has gradually become a controversial issue [26,27], hindering development of Si laser.

2. Moreover, other works were devoted to projecting separated energy levels in NCs to bulk energy bands, which can explain the mixing proportion of direct and indirect transitions. By extracting characteristic points of Kohn-Sham (KS) wavefunctions, a bulk-like energy band was reconstructed. [28–30] However, wavefunction information is lost in the reconstruction process, and the surface state effect is difficult to calibrate, limiting the universality of this method to NCs. In addition, a specific transition can also be roughly calculated by using Kohn-Sham wavefunctions at the highest occupied molecular orbital (HOMO) and lowest unoccupied molecular orbital (LUMO) in real space or $k$-space; thus, a conclusion that Si NCs present a non-phonon assistance quasi-direct optical transition was proposed because of the mixing of **X** and $\Gamma$. [31,32] Nevertheless, this "hybridization" is not intuitive and may not be physical. Because the periodic lattice is broken, it is impossible to avoid the appearance of the separated energy levels observed in experiments [33], following energy band weakening. Thus far, these methods cannot clearly explain whether Si can emit strongly.

More fundamentally, confined-Si emission has been observed since the 1980s [34], but its luminous mechanism remains a mystery. [35,36] One can only roughly borrow momentum dispersion caused by QC to explain why confined Si NCs can emit light but bulk Si cannot.

In this work, we aimed to resolve all of the above problems, and a new physical luminescence model was proposed based on electronic structure analysis in ground-state and excited-state DFT frameworks. Based on the three-dimensional confinement properties of NCs, the specific energy level was discussed in the molecular cluster framework. Since the NCs electronic structure is fundamentally calculated, the key information that has been ignored under the approximate calculation of bulk materials [37], such as QC, will be retained, and the more accurate wavefunction information can be essentially obtained. Here, an intrinsic phonon-free optical transition in Si NCs was demonstrated, addressing the separated energy level of clusters. The transition dipole moment (TDM) with ground-state DFT (GS-DFT) preliminarily determined that Si NCs can achieve a similar direct bandgap transition. More accurate time-dependent DFT (TDDFT) quantitatively analyzed the $\Gamma{\rightarrow}\Gamma$ transition offset and proved that only very small Si NCs can realize a direct bandgap (DBG)-like transition. Our results were contrary to the two previously reported theories, i.e., the redshift of $\Gamma{\rightarrow}\Gamma$ transition and a quasi-direct bandgap with $\Gamma$-$X$ mixing. By extracting the configuration coefficients of excited states, the spatial distribution of electron-hole (e-h) pairs in a single excited phonon-free optical transition was reconstructed. In contrast to the minimal overlap of e-h pairs in bulk Si, QC makes delicate localization of electrons and holes into the core, inducing spatial distortion of excited electrons and overlap of e-h pairs, which strengthens luminescence in Si NCs. This effect is positively correlated with the localized degree. In this work, we named localized state-induced enhanced emission (LIEE) to explain the anomalous luminescence—a phase transition from weak (or none) luminescent state to strong optical transition.

## II. COMPUTATIONAL DETAILS

### A. Ground-state DFT calculation

QUANTUM ESPRESSO [38,39] was employed to obtain all of bulk Si properties with fully-relativistic norm-conserving pseudopotentials. [40] All of based on primitive cell with 60 Ry kinetic energy cutoff and 300 Ry charge density cutoff with PBE functional [41] were chosen to obtain stable structure with 6×6×6 no-shifting auto-generated $k$-mesh. Afterwards, Functional was changed to M06-L [42] to generate a stable self-consistent Kohn-Sham wavefunction. Since the orbital dependence of Meta-GGA functional, 50 energy bands were considered to produce more accurate values. Electron convergence threshold for self-consistency was reduced to $1e^{-9}$ Bohr. The geometric convergence condition was the default.

All GS-DFT calculations of NCs were performed in ORCA 5. [43] Firstly, structures were optimized with PBE [41] functional and triple-zeta def2-TZVP [44] basis set with RI-Coulomb def2/J fitting. [45] Except that SCF was employed with tight convergence, other convergence conditions were adopted with default. The optimized structures were recalculated by different functional at the same computational level, and molden files recording the molecular orbital information were obtained for pse-TDM and the simplified TDDFT (sTDDFT) calculation.

### B. Pse-TDM calculated from molecular orbital

According to the time-dependent perturbation theory, the absorption transition rate is proportional to

$$W_{pq} \sim |<p|\mathbf{P} \cdot \mathbf{e^{ik \cdot r}} |q>|^2 \tag{1}$$

Considering the absorption (or radiation) energy is larger than the size of NC, it is approximated by electric dipole moment $\mathbf{e^{ik \cdot r}} = 1 + i\mathbf{k} \cdot \mathbf{r} + \ldots\ldots$, then

$$W_{pq} \sim |<p|e\,\mathbf{r}\,|q>|^2 \tag{2}$$

The p-th Kohn-Sham (KS) molecular orbital (MO) can be expanded by atomic orbital (AO)

$$\Psi_p = \Sigma_m C_{pm} \psi_m \tag{3}$$

$C_{pm}$ is MO expansion coefficient. For Gaussian-type orbitals (GTO)，the transition dipole moment (TDM) is

$$<p|\,\mathbf{r}\,|q> = \Sigma_m \Sigma_n C_{pm} C_{qn} \Sigma_i \Sigma_j d_{mi} d_{nj} <\phi_i^{\mathrm{GTO}}|\,\mathbf{r}\,|\phi_j^{\mathrm{GTO}}> \tag{4}$$

$\phi^{\mathrm{GTO}}$ is Gaussian-type basis function. $d_{mi}$ and $d_{nj}$ are combination coefficient.

## C. Excited state calculation with (s)TDDFT and wavefunction post-processing

All TDDFT calculations were performed in ORCA.[43] We employed PBE0 functional with RIJCOSX approximation for Coulomb and HF exchange. It must be noted that ORCA turns on Tamm-Dancoff approximation (TDA) by default (off by 'TDA FALSE'). Currently, ORCA only outputs excited states and orbital contributions, but cannot print configuration coefficients for full TDDFT. We extracted all configuration coefficients in CIS file, and then performed excited state analysis with modified Multiwfn.[46] sTDDFT was directly calculated by stda program developed by Grimme et al. [47] We modified stda program to print all configuration coefficients for hole and electron analysis.

## D. E-h pair real space analysis and Coulomb attractive energy

In TDDFT framework, all KS orbital are involved in electron excited state. The excited state wavefunction is

$$\Psi_{\mathrm{TD}} = \sum_{i \rightarrow a} \omega_i^a \Phi_i^a + \sum_{a \rightarrow i} \omega_i^{'a} \Phi_i^a \tag{5}$$

$\Phi_i^a$ is the configuration state wavefunction corresponding to moving an electron from originally occupied i-th MO to virtual a-th MO. $\omega_i^a$ and $\omega_i^{'a}$ are configuration coefficient of excitation and de-excitation, respectively, limited by normalizing conditions. Density distribution of hole and electron can be perfectly defined as [48]

$$\rho_h(\mathbf{r}) = \sum_{i \to a} \left(\omega_i^a\right)^2 \varphi_i(\mathbf{r})\varphi_i(\mathbf{r}) + \sum_{i \to a} \sum_{i \neq j \to a} \omega_i^a \omega_j^a \varphi_i(\mathbf{r})\varphi_j(\mathbf{r}) \tag{6}$$

$$\rho_e(\mathbf{r}) = \sum_{i \to a} \left(\omega_i^a\right)^2 \varphi_a(\mathbf{r})\varphi_a(\mathbf{r}) + \sum_{i \to a} \sum_{i \to b \neq a} \omega_i^a \omega_i^b \varphi_a(\mathbf{r})\varphi_b(\mathbf{r}) \tag{7}$$

$\varphi_i$ and $\varphi_j$ are occupied molecular orbitals, while $\varphi_a$ and $\varphi_b$ are virtual orbitals. Thus, e-h pair Coulomb attractive energy can be calculated via simple Coulomb formula:

$$E_c = \iint \frac{\rho_h(\mathbf{r_1})\rho_e(\mathbf{r_2})}{|\mathbf{r_1}-\mathbf{r_2}|} \mathrm{d}\mathbf{r_1}\mathbf{d\mathbf{r_2}} \tag{8}$$

whose negative value is known as exciton binding energy.

## III. RESULTS AND DISCUSSIONS

### A. Γ→Γ shift and enhanced optical transition with GS-DFT pseudo-transition

Despite the limited description of unoccupied orbits in GS-DFT, it can qualitatively obtain partial transition information, which is helpful for understanding the optical transition in Si NCs, especially the controversial issues of whether direct-transition redshift with size reduction is reasonable [25–27] and whether Si NCs can achieve direct transition.[19,24,49] A simple physical model assumes that an optical transition occurs in a pair of orbitals obtained by GS-DFT, that is, considering that each excited state has only one pair of orbital contributions. When excited states are linearly combined by multiple GS-DFT orbitals, this corresponds to the description in TDDFT. According to time-dependent perturbation theory, when the initial and final state wavefunctions are extremely accurate, all transition information can be obtained theoretically. However, accurate acquisition is very difficult, especially for large systems with a few hundred atoms. DFT neglects the multielectron problem and can be used to simulate very large systems. In the GS-DFT frame, the transition dipole moment (TDM) is defined as the pseudo-TDM (pse-TDM), which represents the transition probability from the initial state to the final state. Through numerical calculation, pse-TDM was easily obtained in all transition states (see Methods).

Size-varying Si NCs were calculated, and the pse-TDM diagram is shown in **Fig. 1a**. When the particle size is greater than 1.2 nm (Si$_{130}$ cluster), similar to bulk silicon materials, Si NCs show a long absorption tail, which proves that Si NCs inherit the characteristics of an indirect bandgap (IBG). It should be emphasized that the concept of the energy band is neglected in our calculations. Therefore, this "IBG" is attributed to the very low TDM near the gap energy. More broadly, all NCs with long absorption tails, i.e., very low transition dipole moments, can be classified as IBG-like materials. Nevertheless, when the size of the Si NC is less than 1.2 nm, the peak shifts slowly, gradually closing to the gap energy. A strong band-edge peak with DBG-like characteristics appears around $Si_{30}$. We define this small spike as the "edge" transition. As the size decreases, the "edge" transition coincides with the $\Gamma{\to}X$-like transition (**Fig. 1b**), which means that Si NCs can achieve a strong transition process near the optical gap, i.e., a DBG-like transition. The orbital corresponding to the "edge" transition comes between the frontier HOMO ($\Gamma$-point energy edge) and a deeper LUMO (**Table S1**). We attribute this transition to increased overlap between initial and final DFT wavefunctions with size reduction, possessing electron localized properties in the molecular orbital isosurface **(Fig. S4)**, corresponding to momentum dispersion caused by the QC effect in energy band theory (**Fig. 1c**).

Another critical excited state is the $\Gamma{\to}\Gamma$ optical transition. In bulk Si, experiments [50] have shown that the absorption peak of 3.4 eV mainly comes from the band-edge direct transition, including the $\Gamma{\to}\Gamma$ transition. Considering that holes tend to relax to the top of the valence band in the radiative recombination process, the direct band-edge transition on the Brillouin path other than the $\Gamma{\to}\Gamma$ transition is not the main concern. According to the relative position of the absorption peak and the corresponding transition orbital number (**Table S2**), the absorption peak corresponding to the $\Gamma{\to}\Gamma$-like transition is calibrated (**Fig. 1a** and **1b**). With the decrease in nanocrystalline size, contrary to previous reports, the $\Gamma{\to}\Gamma$-like transition energy gradually increases and flattens when the size is less than 0.8 nm. In fact, in $Si_{22}$ clusters, the $\Gamma{\to}\Gamma$ transition energy is 0.23 eV higher than the $\Gamma{\to}X$ transition energy, which is much higher than the thermal kinetic energy of electrons at room temperature and cannot emit light stably from the $\Gamma$-point without relaxation to the lowest $X$ state. Nevertheless, a strong transition may show DBG characteristics when the particle size is below 0.8 nm due to a tiny difference between the "edge" and $\Gamma{\to}X$ (**Fig. 1b**). When hot electrons relax on the excited state energy surface, this slight gap leads to a shift of the luminescence center from the $X$-point to the "edge" state due to the strong transition of the "edge". In addition, the smaller the size is, the stronger the confinement and the greater momentum deviation allowed for a high-probability transition (**Fig. 1c**). To verify the conclusion with a different functional, B3LYP functional [51] was used and showed a similar trend (**Fig. S5**). A more concrete discussion is given below under a more precise TDDFT framework.

## B. Optical transition with (s)TDDFT excited states calculation

To determine the $\Gamma\rightarrow\Gamma$-like transition shift and whether a DBG-like transition can realize size reduction more accurately, a more rigorous theory must be adopted. Full TDDFT has been very successful in predicting the photoelectric properties of many clusters. [52–57] In previous reports, only excited states of very small Si clusters ($< Si_{40}$) were calculated [58,59], and the above problems were not considered. Here, NCs up to $Si_{244}$ were calculated to gain insight into the optical transition in Si NCs. NCs smaller than 1.2 nm (~$Si_{82}$) were studied within up to 200 excited roots, while other NCs were limited to just up to the second excited state peak (**ES peak 2**) by extremely expensive computational costs. The phonon-free optical transition in Si NCs is shown in **Fig. 2a**. All transitions show a first excited state peak (**ES peak 1**) that differs from the long absorption edge in bulk silicon. With a decrease in NC size, the oscillator strength of **ES peak 1** gradually increases from $8\times10^{-4}$ to $1\times10^{-2}$. The energy of **ES peak 2** is slightly higher than **ES peak 1** by 0.3 eV, and its oscillator strength is 3-10 times larger.

The simplified TDDFT (sTDDFT) proposed by Grimme *et al.* [47] can be used to calculate ten thousand excited states much more economically than full TDDFT. Aiming at the overall transition in Si NCs, a spectrum was calculated up to 14 eV using sTDDFT. For a comparison of TDDFT and sTDDFT spectra, refer to **Fig. S6**. The sTDDFT excited states have a slight redshift (0.04-0.4 eV) and an oscillator strength that is almost the same order of magnitude. Overall, sTDDFT can better ensure consistency with TDDFT despite the partial loss of excitation energy and oscillator strength. All optical absorption spectra are similar to that of bulk silicon (**Fig. 2b**), showing wide envelope behavior at different photon energies. Obviously, the smaller the system is, the wider the envelope (bulk-Si vs. $Si_{22}$, 2 eV vs. 5 eV). This behavior is inevitable after the electron energy level is coupled with size reduction, resulting in the original compact energy band being divided into separate energy levels (**Fig. S3**).

The $\Gamma\rightarrow\Gamma$-like transition always occurs from the uppermost HOMO to LUMO-n. Thus, the $\Gamma\rightarrow\Gamma$-like transition can be identified by inspecting the orbital contribution of all excited states. Because the $\Gamma\rightarrow\Gamma$ transition in bulk silicon includes two parts with different energies, the low-energy transition can be denoted by $\Gamma\rightarrow\Gamma_1$, and the high-energy transition is written as $\Gamma\rightarrow\Gamma_2$ (**Fig. S1**). Their energy difference represents the envelope width of the absorption spectrum. All $\Gamma\rightarrow\Gamma$-like transitions were derived from HOMO (0,1,3)-> LUMO-n (**Table S3 and Table S4**). Moreover, atomic orbitals can more directly reflect the type of transition. By printing all main contributions to one excited state, Atomic orbital components in different excited states were obtained. In bulk Si, $\Gamma \rightarrow \Gamma_1$ is often composed of a P-shell to P-shell transition, while $\Gamma \rightarrow \Gamma_2$ corresponds to a P-shell to hybrid SP-shell transition. A concordant result was found in Si NCs (**Table S5**). All $\Gamma \rightarrow \Gamma_1$-like transitions are between P-shells, and all $\Gamma \rightarrow \Gamma_2$-like transitions are P-shell to hybrid SP-shell. Contrary to a previous report, it is observed that the $\Gamma \rightarrow \Gamma$-like transition energy increases as the NC size decreases and never approaches the $\Gamma \rightarrow X$-like transition, which means that the previous theory about realizing a strong direct transition caused by the coincidence of the $\Gamma \rightarrow \Gamma$ and $\Gamma \rightarrow X$-like transitions with Si NC size reduction [24] is flawed, as shown in **Fig. 2c**. Notably, energy level separation or interface capture is likely to have an unsteady rapid transition, as reported in the literature [19,25,60], resulting in a redshift phenomenon that cannot be attributed to the $\Gamma \rightarrow \Gamma$ transition. Another statement about the mixing of **X** and $\Gamma$ [49] remains to be discussed because of a missing observation of strong $\Gamma$-**X** coupling coming from the atomic energy level or $\Gamma$-point redshift in our basic results. Similar to GS-DFT, the optical transition indeed differs from the long absorption edge in bulk Si. Especially in $Si_{22}$, the oscillator strength can reach approximately 0.015 (**Fig. 2a**). When the NC size increases to 2 nm, the oscillator strength sharply decreases to only $10^{-6}$. This behavior is also the reason most orange or red fluorescent Si NCs always show more than a microsecond fluorescent lifetime, while that of blue can only reach nanoseconds.

The radiative recombination lifetime of various-sized Si NCs can be estimated based on the present results. Stable fluorescence should be produced from the lowest excited states, while **ES peak 1** has typical emission attributes, i.e., the lowest state and a large oscillator strength. As the particle size increases, the radiative recombination lifetime increases exponentially (**Fig. 2d**). A microsecond lifetime is obtained when the NC size is approximately 1.5 nm. Significantly, $Si_{22}$ intrinsically possesses an ~60 ns radiative recombination lifetime, which is slightly slower compared with $CsPbBr_3$ NCs.[37] The above results show that small Si NCs may have a fast intrinsic DBG-like (but not $\Gamma \rightarrow \Gamma$-like or mixing $\Gamma$-**X**-like) transition, while a slightly large size may lead to a slow IBG-like (tend to $\Gamma \rightarrow X$-like) transition.

### C. Localized state induced enhanced emission caused by QC

According to the above analysis, **ES peak 1** plays an important role in the anomalous luminescence of Si NCs, much different from the long absorption edge in bulk Si. A full understanding of this difference will help advance luminescence physics. In bulk Si, the conduction band maximum (CBM) and valence band minimum (VBM) are offset in $k$-space, which inhibits emission of bulk Si. Before the separated electronic structure forms an energy band, the arrangement of the electron cloud is affected by interface scattering, thus breaking the perfect Bloch wave in a periodic system. This "break" is likely to cause an electron shift and realize abnormal luminescent enhancement in Si NCs.

Considering the configuration interaction singles (CIS) attribute of TDDFT, all excitations can be described as an electron deviating from a hole and turning into a higher energy electron. The contribution of each silicon atom in the NCs to the transition matrix can be easily investigated from the basis function in real space by Multiwfn. [46,48] All **ES peaks 1** and **2** show local excitation characteristics, and the proportion of Si atoms participating in the optical transition decreases with increasing size (**Fig. S7**). More specifically, electrons and holes are localized in the NC core (**Fig. S8**). In bulk Si, the **Γ** VBM is distributed between silicon atoms, including two directions (**Fig. S9**), while CBM close to **X** presents the form of bisecting the angle between two silicon bonds, which also bisects the angle between the two directions of CBM in another plane. This minimal overlap results in emission from bulk Si being very difficult, also known as momentum mismatch. However, this minimal-overlap situation is changed in the NCs (**Fig. 3**). The hole distribution is very similar to that of the VBM in bulk silicon, showing Si-Si bonding characteristics, while the antibonding CBM (electrons) is changed with local migration caused by boundary scattering, resulting in an electron cloud deviating from the original bisector dihedral state (**Fig. 3f**), thus enabling partial overlap of electrons and holes.

Another important change is that truncated Bloch waves concentrated in the center of NCs also provide additional overlap. For spherical Si NCs, $sp^3$ hybridization makes the localized wave located inside the tetrahedron and extend towards four vertices (**Fig. 3g**). More subtly, contrary to the situation of bisection to silicon bond-bond angle in bulk Si, vertices of the local wave are tetrahedrally arranged with ten silicon atoms, contributing to partial overlap of the e-h pair. These two overlapping effects jointly lead to the free-phonon-assisted optical transition in Si NCs. Compared with the latter effect, the former is mainly concentrated at the edge of the local state. Therefore, we distinguish these effects by the localized state edge and center.

Localized state edge and center are also applicable to the analysis of extremely small Si clusters. Here, two special examples include only one of these two effects, corresponding to a localized state center in $Si_{10}$ and a localized state edge in $Si_{17}$. Different from the previous conclusion of **Fig. 2d** that the oscillator strength of **ES peak 1** increases as the size decreases, that of $Si_{10}$ decreases (**Fig. 4**). In $Si_{10}$, the first excited state exhibits transition prohibition due to a symmetry restriction (**Fig. 4c**, the inset of **Fig. 4a**), while **ES peak 1** and **2** are contributed from the localized state center (**Fig. 4d** and **4e**). Compared with $Si_{22}$, the oscillator strength of **ES peak 1** and **2** was reduced 10-fold because of a limited effect of localized states centered on the enhanced optical transition, which also reflects that the localized state edge plays an essential role in the optical transition of Si NCs. Nevertheless, the oscillator strength of **ES peak 1** can still be compared with that of $Si_{66}$ (~0.002), which is much higher than that of IBG-Si. In $Si_{17}$, in addition to the localized state edge, a partial electron wave is also trapped near the central Si atom, proving that the truncated Bloch wave would be localized not only in a tetrahedron with ten silicon atoms tending to form in larger clusters (**Fig. 3g**) but also on the central silicon atom with a superlocal state (**Fig. 4f** and **4g**). Thus far, both in large NCs and small clusters, the intrinsic phonon-free optical transition in Si NCs can be attributed to the constraint of the e-h pair, i.e., the localized state makes Si (NCs) emit light possibly.

With increasing size, the truncated Bloch wave expands, and the e-h pair becomes more diffuse in real space, which can be characterized by e-h Coulomb attractive energy (**Fig. 5**). In Si NCs, the oscillator strength is strictly positively correlated with the e-h Coulomb attractive energy. Localized states in **ES peak 1** and **2** have an extremely high exciton Coulomb energy, up to 3.7 eV for small NCs, which is much higher than some e-h delocalized materials. [61] Even for large NCs with a microsecond radiative lifetime, this value still reaches approximately 2 eV. For the optical transition in the same system, the localized degree can be calibrated by e-h Coulomb attractive energy. The larger the exciton Coulomb attractive energy is, the higher the localized degree. A higher localized degree often enables a greater offset of the electron wavefunction, making the above two overlapping effects more obvious. This result is the most fundamental reason a fast optical transition has been observed in small Si NCs in experiments [62–64] and calculations.

## IV. CONCLUSION

In summary, we use results from GS-DFT and (s)TDDFT to discuss intrinsic phonon-free optical transitions in Si NCs. Firstly, we reported the $\Gamma{\rightarrow}\Gamma$ transition is blueshifted and gradually tends to be stable, contrary to previous reports on the redshift of $\Gamma{\rightarrow}\Gamma$ transition with size reduction resulting in fast direct-optical-transition realization and **Γ-X** mixing quasi-direct transitions bringing about luminescence of Si NCs. Most importantly, based on the principle of exciton luminescence, the spatial distribution of electrons and holes in NCs with different sizes was shown. Compared with the traditional understanding of luminescence in bulk silicon, that is, momentum dispersion-induced emission, localized state-induced enhanced emission caused by QC was proposed. In Si NCs, QC bends the electron wavefunction, thus breaking the minimal overlap condition of electrons and holes in bulk Si. At the same time, the directivity of the electron wavefunction localized in the core also makes a positive contribution to the optical transition. This enhanced optical transition is largely affected by the NC size. The more localized the degree is, the greater the oscillator strength, which is positively correlated with the exciton binding energy. LIEE has supplemented luminescence physics about long-standing luminescent cognition, and would promote the development of fast optical transitions in Si NCs and other luminescent materials, and can be expected to be used in the design of lasers, LEDs, etc.

## AUTHOR CONTRIBUTIONS

Conceptualization, Data curation, Investigation, Methodology, Software, Visualization, Writing – original draft: F.W.; Funding acquisition, Project administration, Supervision: S.Z.; Resources: S.Z. and Q.O.; Writing – review & editing: F.W. and S.Z..

## ACKNOWLEDGMENTS

This work is supported by Science and Technology commission of Shanghai Municipality (21ZR1408800), and National Natural Science Foundation of China (11975081).

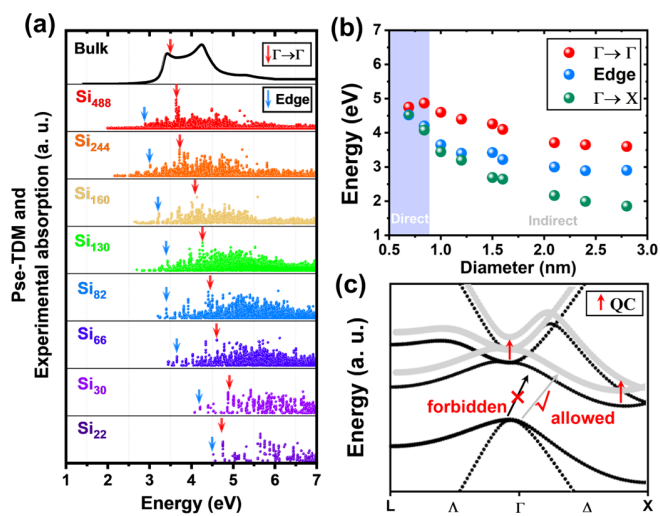

**Fig. 1.** Pse-TDM obtained with M06-L functional shows (a) **Γ→Γ** (red arrow) and "edge" transition, (b) that change versus particle size. The **Γ→X**-like energy represents energy gap in Si NCs. **Γ→Γ** was determined through the transition orbital number of HOMO to LUMO-n (**Table S2**) with a strong transition peak. The blue shadow indicates that a direct transition may occur with "edge" excitation peak so close to energy gap in GS-DFT. (c) Enhanced quantum confinement is responsible for **Γ→Γ**, **Γ→X** blue shifts and allowable optical transition achieved by momentum dispersion.

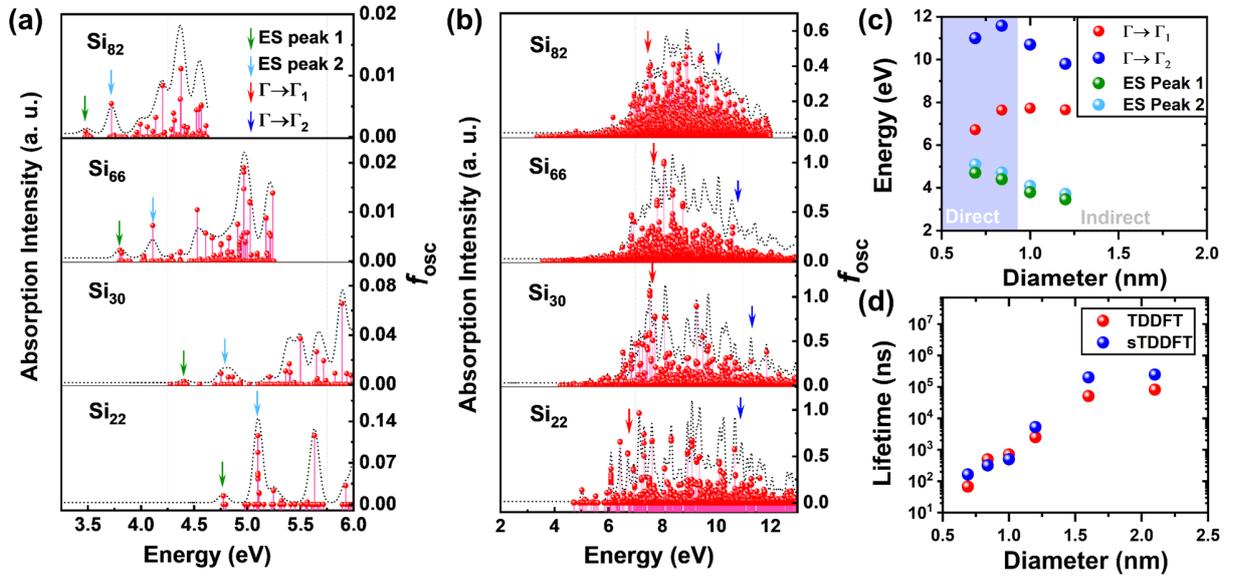

**Fig. 2.** Excited state calculation in Si NCs. Absorption spectrum (all broadened by Gaussian profile with 0.1 eV) and oscillator strength obtained by (a) TDDFT and (b) sTDDFT. (c) Some characteristic peaks move with size. (d) Radiative recombination lifetime versus size in Si NCs under full TDDFT and sTDDFT. Red and blue rows were identified by inspecting the orbital contribution of all excited states because $\Gamma \rightarrow \Gamma_1$ is often composed of a P-shell to P-shell transition, while $\Gamma \rightarrow \Gamma_2$ corresponds to a P-shell to hybrid SP-shell transition (**Table S5**).

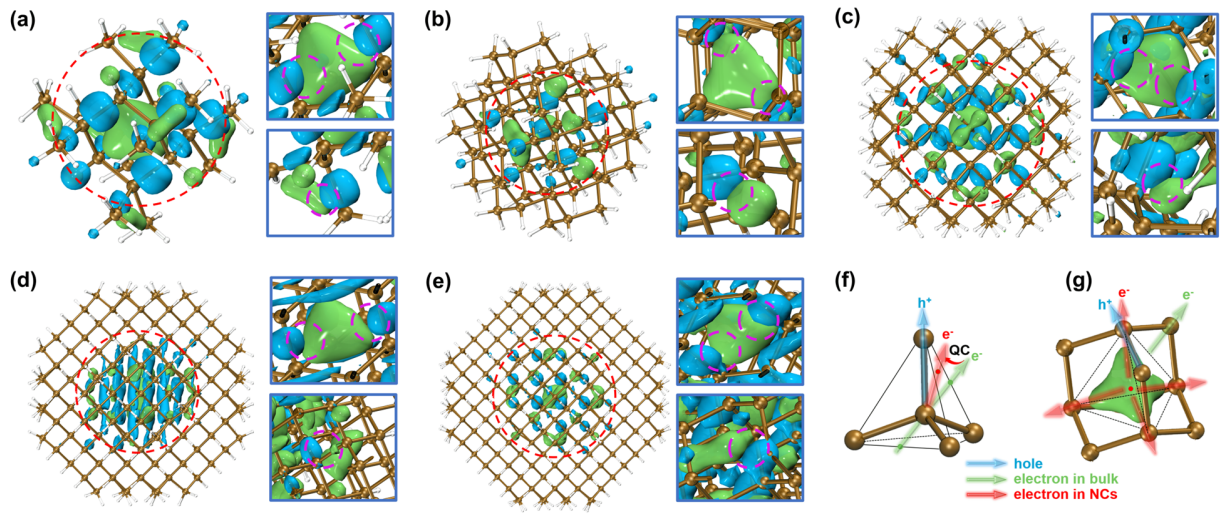

**Fig. 3.** QC induced localized states that allow optical transition with two overlapping parts, (a) for Si$_{22}$, (b) for Si$_{66}$, (c) for Si$_{130}$, (d) for Si$_{244}$, (e) for Si$_{488}$. Blue and green isosurface represent hole and electron, respectively. The upper right of each graph contains localized state center, while the lower contains the localized state closing to edge. Dashed circles mark overlapping areas. These phenomena are described by simple strokes with hole and electron spatial distribution in (f) localized state edge and (g) localized state center.

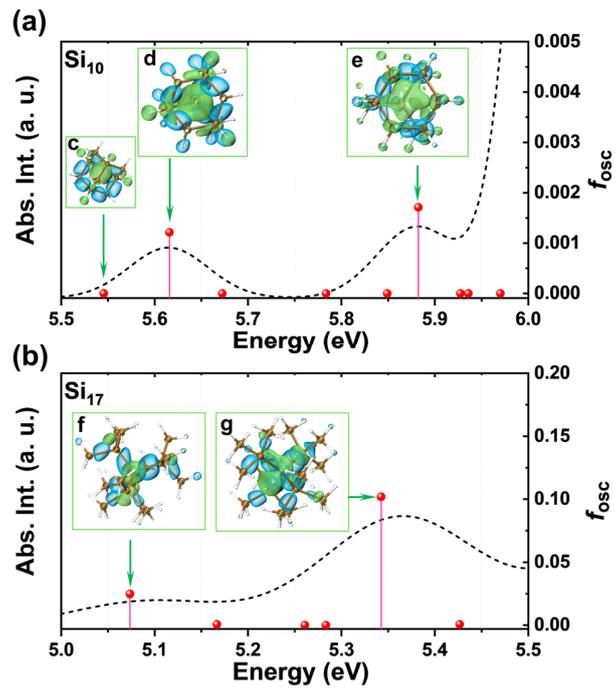

**Fig. 4.** Optical transition in much small Si clusters, showing a pure localized state center in (a) $Si_{10}$ and a special situation in (b) $Si_{17}$, with spatial distribution of e-h pair (Blue and green isosurface represent hole and electron, respectively.) both in **ES peak 1** and **2** (c-g).

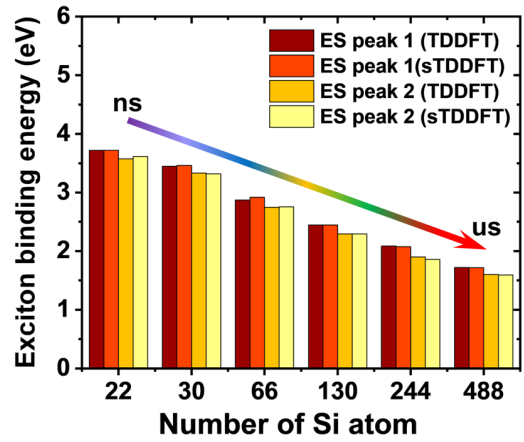

**Fig. 5.** Exciton binding energy from e-h Coulomb attractive energy in two localized states with

(s)TDDFT. There is little difference (max error 2%) in results of sTDDFT and TDDFT.